# White Paper:
# Integrated Circuit Design in US High Energy Physics
### July 10, 2013

Editor: G. Haller (SLAC)

Authors:

G. De Geronimo (BNL)

D. Christian (FNAL)

C. Bebek, M. Garcia-Sciveres, H. Von der Lippe (LBNL)

G. Haller (SLAC)

A.A. Grillo (UCSC)

M. Newcomer (UPenn)

# Table of Content

# 1 Executive summary

Custom "Application Specific Integrated Circuits (ASICs)" were first developed for HEP in the 1980s to read out silicon strip vertex detectors in colliding beam experiments. They have since become one of the core technologies available to detector designers.

A number of factors make ASICs essential to HEP. These include:
- Small physical size: The space constraints of many detectors, most notably pixel vertex detectors, absolutely require custom microelectronics. Even when commercial electronics can be used, small ASICs can often be positioned closer to the sensor than would otherwise be possible. This reduces input capacitance and improves noise performance. In many cases the size of the cable plant is also reduced (e.g. smaller radiation length, less power dissipation on the detector, lower cooling)
- Low power dissipation: The infrastructure required to power and cool on-detector electronics often limits detector performance. Especially in high channel count applications, low power dissipation can be one of the most important specifications.
- Radiation tolerance: Many HEP applications require 10 – 100 Mrad Total Ionization Dose (TID) tolerance as well as immunity against Single Event Upsets (SEUs). Future vertex detectors may require Grad tolerance.

ASIC-related R&D is required in a number of areas in order to improve science output or simply make possible future experiments in the intensity, cosmic, and energy frontiers. Examples are:
- high-speed waveform sampling
- pico-second timing
- low-noise high-dynamic-range amplification and pulse shaping
- digitization and digital data processing
- high-rate radiation tolerant data transmission
- low temperature operation
- extreme radiation tolerance
- low radioactivity processes for ASICs
- low power
- 2.5D and 3D assemblies

A workshop was held on May 30 to June 1, 2013 to discuss and edit this document with input from the US HEP IC design community. Findings are given to summarize the major points from the workshop, see section 5.11. Based on those findings, several recommendations are made for furthering HEP IC design activities in the US, see section 5.12.

# 2 Introduction

## 2.1 Motivation and scope

This whitepaper summarizes the status, plans, and challenges in the area of integrated circuit design in the United States for future High Energy Physics (HEP) experiments. It has been submitted to CPAD (Coordinating Panel for Advanced Detectors) and the HEP Community Summer Study 2013 (Snowmass on the Mississippi) held in Minnesota July 29 – August 6, 2013.

A workshop titled "US Workshop on IC Design for High Energy Physics – HEPIC2013" was held May 30 to June 1, 2013 at Lawrence Berkeley National Laboratory (LBNL). The motivation, agenda, presentations, and list of attendees are posted at the following location: https://indico.physics.lbl.gov/indico/conferenceDisplay.py?confId=2. A draft of the whitepaper was distributed to the attendees before the workshop, the content was discussed at the meeting, and this document is the resulting final product.

The scope of the whitepaper includes the following topics:
- Needs for IC technologies to enable future experiments in the three HEP frontiers – Energy, Cosmic and Intensity Frontiers
- Challenges in the different technology and circuit design areas and the related R&D needs
- Motivation for using different fabrication technologies
- Outlook of future technologies including 2.5D and 3D
- Survey of IC's used in current experiments and IC's targeted for approved or proposed experiments.
- IC design at US institutes and recommendations for collaboration in the future.

## 2.2 Importance of IC design for HEP

Custom integrated circuits, usually referred to as ASICs (Application Specific Integrated Circuits), were first developed for HEP in the 1980s to read out silicon strip vertex detectors. This miniaturization of SSD readout electronics was crucial to the successful development of vertex detectors for colliding beam experiments. Subsequent advances in vertex sensor technology, most notably hybrid silicon pixel detectors, have been tightly coupled to advances in ASIC functionality.

Although ASICs were first used in vertex detectors, virtually all large scale detector readout systems have seen significant benefit from the use of ASICs. Custom integration allows higher density, enhanced circuit performance, lower power consumption, lower mass, much greater radiation tolerance, or low temperature performance than is possible with commercial ICs or discrete components. In some applications even low channel or transistor count require the use of ASICs to enable better science; examples are several ASICs for space and for on-ground detector applications. Analog memories have replaced delay cables in many experiments. In others, mixed analog/digital ASICs have facilitated signal digitization and digital storage close to the detector. ASICs (such as content addressable memories) have also been used as key elements of trigger systems. Integrated circuit technology has improved with breathtaking speed for decades, and will continue to do so. The applications of new integrated circuit technologies will likely lead to transformative detector developments that enable future experiments in the same way that miniaturization enabled vertex detectors for colliding beam experiments.

### *2.3 ASICs for other applications*

Nowadays integrated circuits are used in numerous applications in a wide variety of fields. The importance of ASICs in enabling science is also true for other fields as e.g. photon sciences or nuclear physics.

The IC design groups at all HEP DOE institutes supply ICs for non-HEP areas, in fact for many institutes the HEP work is a fraction of the total effort by the respective design groups. The fraction may vary across institutes and throughout the years. Currently the minimum percentage HEP related work required to sustain the group sizes varies between 10% and 90% depending on the institute. A large fraction of the non-HEP IC's designed by DOE HEP laboratories is for x-ray and neutron imaging, scattering, and spectroscopy experiments, mainly under DOE BES. A smaller fraction is for space, medical imaging, national security, and industrial applications (under CRADA or WFO programs). This diversification allows the IC design groups to maintain critical size, keep pace with the rapid progress of CMOS technologies, and maintain the circuit designs at the level of state-of-the art. Without diversification the US IC community wouldn't be able to efficiently respond to the HEP needs. Although the requirements of the various applications might differ, the circuit design methodologies are similar. Additionally, the CAD (Computer-Aided Design) tools are the same, resulting in lower effective cost for HEP due to cost sharing and also allow minimizing labor overhead because of a more efficient management of engineering time to level peaks and valleys in HEP work. The number of ASIC designers at DOE laboratories (BNL, FNAL, LBNL, and SLAC) is between 4 and 6.

There are also examples where ASICs developed for HEP are used in experiments in other fields, as they are or with some modifications, since there is an overlap of detector technologies between different fields. Examples are ASICs for RHIC and CBAF for Nuclear Physics.

### *2.4 Collaborations*

Most ASICs in current and proposed HEP experiments (and that is also true for nuclear physics and photon science experiments) are designed entirely at single institutes. One of the reasons is that they are not complex enough to warrant the partitioning of the design into smaller blocks, which are worked on at several institutes. When the design is distributed because it is too complex for one institute, there is overhead to coordinate circuit and layout interfaces and functionality. Typically one institute coordinates the design and integrates the sub-blocks.

Even for an ASIC designed entirely at one institute, it can be helpful to exchange with other groups prototype test results (for example on radiation effects), circuit blocks, and even actual prototypes. In general, circuits from a particular ASIC cannot be used unchanged for another application even if the ASIC technology is exactly the same, and often the effort to modify an existing circuit design is the same as to design a new one.

However, through collaboration, the experience and lessons learned in one ASIC can and should be applied to the design of another. Direct communication between designers is essential for this to happen.

For some of the LHC ASIC's the effort required was too large for any single institute. As an example, the FE-I4 pixel readout chip recently developed for ATLAS was designed by a collaboration of 12 designers working at 5 institutes and one commercial vendor, in 5 countries. Layout, simulation, and schematic design were done at most of the institutes. The full design library with all views was shared in a repository available at all of the institutes, just as is done for software development. This type of collaboration requires coordination of CAD/CAE tools and management of proprietary material, as well as good communication.

The design and verification of smaller feature size processes (65nm and below) are getting more complex, and another area of collaboration could be the sharing of knowledge or even partitioning between institutes on the individual design or verification steps.

# 3 Role and organization of DOE HEP laboratories and universities in ASIC design

### 3.1.1 Role of US HEP IC design groups

In general the role of US HEP design groups is to
- work with scientists to find out what is possible for experiments (functionality, performance, location, space, power, etc)
- work with scientists and engineers to design ASICs to optimize integrated systems from sensor to DAQ (mechanical, electrical, cooling)
- provide ASICs for approved experiments (design, test stand-alone, test in detector system)
- perform targeted R&D for future candidate experiments
- perform generic R&D to advance state of art for HEP ASICs
- maintain HEP-specific expertise and keep up with ASIC technology (requires stable work-force since experts in this area can't be let go and hired at will)

### 3.1.2 Distributed ASIC design capabilities

In the following section, advantages and disadvantages of having ASIC design capability at several US institutes are discussed.

Reasons for distributed design capabilities:
- Engineer-physicist interaction: A tight interaction between engineer and physicist enables a better performing sensor/ASIC system, typically the ASIC is designed at the institute where the detector physicist is located.
- System Issue: Most IC's targeted for future experiments are embedded very tightly in the detector system and therefore are best designed at the institute providing the respective system. This is a very important point especially for front-end ASICs located close to the sensors.

- Innovation: As in the case of science, ASIC design capability at different institutes enhances innovation. That is important because ASICs often enable science which would not be possible without ASICs.
- Management of schedule risk: In-house design allows prioritization to reduce schedule risk. Outsourcing the design to another institution risks being assigned a lower priority compared to their in-house efforts, especially if in-house ASICs are delayed due to unexpected design or performance issues. Having capability in-house avoids such inter-laboratory issues.
- ASICs for other fields: HEP ASIC design is only a part of the ASIC design effort undertaken at DOE institutes. Others are for BES, other DOE branches, and non-DOE customers. That benefits all parties since circuit techniques learned can be applied to all fields.

Possible disadvantage of distributed design capabilities:
- Cost: The engineering cost for the ASIC design itself (FTE's) is about the same whether an ASIC is designed at a remote institute or where the detector sub-system work is performed. The fabrication cost for the ASICs is the same since all laboratories already use third-party multi-project organizations for fabrication (e.g. MOSIS or Europractice). The main additional cost is for the CAD tools which can be significant (several hundred thousand dollars/year). Although the total number of licenses is about the same, there is a base cost of obtaining and maintaining the tools. Not every design requires expensive tools; some institutes also use lower cost CAD tools which are satisfactory for most steps in the design and verification process.
- Potential isolation of groups: If the ASIC designer at institute A is not aware of the developments at institute B then there is potentially duplication of the effort and increased development time. This can be mitigated by fostering good communication between groups (see recommendation).

Besides laboratories, US universities have always played an important role in the design, development and implementation of instrumentation for high energy and nuclear physics detectors. This includes the wave of innovations since the late 1980's that have exploited ASICs to enable the development of highly granular large scale detector systems. Twenty of the ASICs listed in the table in the appendix were developed wholly or in part by universities in collaboration with laboratories. As we look to the future, it will be important to retain a resident knowledge of ASIC design development and testing within the university community for training the next generation of physicists and enabling a common level of understanding for innovation and capacity for contribution to large scale future experiments, including knowing when and how to use ASICs in favor of or along with other technologies.

In collaborating with laboratories, universities play an important role in training young physicists to design optimized instrumentation for physics experiments. This clearly includes the understanding the capabilities and limitations of ASICs.

# 4 Future needs for ICs to enable experiments and better science

## *4.1 Energy frontier*

### 4.1.1 ATLAS and CMS

In the near term there is a large need for ASICs for upgrading the ATLAS and CMS experiments. Each experiment currently contains approximately 40 different ASIC designs. Within 10 years the experiments will replace their tracking systems as well as the electronics of many calorimeter and muon detector elements. The number of ASIC designs needed for this will be of a similar scale. Between 10 and 20 ASICs will be designed in the US or in collaborations involving US design groups for ATLAS or CMS over the next 5 to 7 years. Probably most of these ASICs will use the 130nm CMOS process, which is very well understood, will remain available for a number of years, and is accessible via a CERN frame contract with a foundry. However, some ASICs will need more advanced technologies (e.g. 65nm CMOS, SOI, or SiGe), mainly to achieve the maximum possible logic density and/or data transmission speed. It is also possible that monolithic active pixels (MAPS) will make enough progress towards radiation hardness in the near future that they will be used in LHC detectors. The main areas where challenging designs are needed are:

- Hybrid pixel detector readout
- Radiation tolerant high-rate MAPS
- Coupled layer "intelligent" tracking detectors
- Associative memories for fast track finding hardware
- Radiation tolerant high speed data transmission (includes calorimeter and tracker readout and timing, trigger, and control signal distribution)

The LHC upgrades have by far the single largest ASIC needs for HEP this decade, but these are fairly well understood and therefore not repeated at length in this document. For details see references 1-4 below.

1. CMS Collaboration, "CMS technical proposal: Upgrade of CMS detector through 2020," CERN-LHCC-2011-06 (2013).

2. ATLAS Collaboration, "Letter of Intent for the Phase-II Upgrade of the ATLAS Experiment," CERN-LHCC-2012-022. (2013).

3. F. Anghinolfi, et. al.,"R&D Activities in Electronics for future HEP Experiments ," submitted contribution to European Strategy Group, see CERN-ESG-005 (2013).

4. RD53 Collaboration, "RD Collaboration Proposal: Development of pixel readout integrated circuits for extreme rate and radiation ," CERN-LHCC-2013-008 (2013).

### 4.1.2 Electron linear and muon colliders

In the longer term there will be new detectors for new colliders beyond the LHC. This includes electron linear colliders (ILC and/or CLIC) and eventually a muon collider. For electron linear colliders the main challenges are high precision calorimetry (to differentiate W and Z on an event-by-event basis) and highly granular trackers with extremely low mass (capable of resolving c as well as b vertices). Several innovative techniques and technologies are being developed, such as pulsed power (to permit air cooling) and 3D MAPS in which multiple circuit layers are bonded together and thinned to less than 10 microns per layer. In the past, needs for inner tracker detectors were driving the R&D effort towards 3D assemblies.

Several ASICs are being designed in the US, e.g. KPiX and Bean (see ASIC table in the appendix) applicable for several of the sub-systems. More research is ongoing to achieve the System-On-Chip functionality (e.g. 1,000 channels each with calibration, amplification, shaping, sample selection, analog storage and on-chip 13-bit digitization) while still realizing the required noise performance and low power.

Muon collider detectors will have the same physics requirements (precise calorimetry and highly granular low mass tracking) as electron collider detectors, but they will also need to tolerate a high level of background caused by decays of beam muons. Electrons from muon decays will be bent to the inside of the storage ring and will radiate a large number of photons in the process. Current shielding designs reduce the number of photons seen by a detector by a factor of 500, but a large number of neutrons are produced in the shielding and escape into the detector. With shielding, the total dose delivered to detector elements will be at least a factor of ten less than in the upgraded LHC, but still significant. The instantaneous rate of background hits in the detector will be very high. Very good (ns) timing resolution will be required to suppress these backgrounds.

## 4.2 Intensity frontier

### 4.2.1 Cryogenic detectors

#### 4.2.1.1 LBNE

The Liquid Argon Time Projection Chamber (LAr TPC) is a technology of choice for some experiments, e.g. LBNE. The number of sense wires for a ~33 ktons chamber could be up to a million. Extracting this number of (signal) wires from the cold volume is a major mechanical and cryostat design challenge. In addition, the signal to noise that can be achieved will be poor if the sense wire runs are very long. Therefore, the preferred solution is to place electronics with a high degree of multiplexing (hundreds to thousands) in the cold volume.

The design of ASICs for operation in LAr (~ 70K) poses many challenges. The ASICs will need to provide low-noise charge amplification, filtering, sample-and-hold, analog-to-digital conversion, and digital multiplexing in two or more stages. The ASICs must satisfy requirements of low power dissipation and continuous operation without failures for a long time. Additionally, digital and voltage regulator ASICs might be needed.

#### 4.2.1.2 Very low background experiments

Generation 3 direct dark matter searches will also enter the multi-ton regime. The technology will be different in detail (possibly Xe instead of Ar, dual phase with readout in the gas phase, etc.), but they will contain a much larger number of channels in a cold volume compared to generation 2 experiments. While significantly smaller than LBNE, placing electronics inside the cold volume may still be beneficial to the performance of these detectors. Low radioactive emission requirements prevent the use of most commercial electronics devices inside the detector volume. It has been shown (i.e. for EXO) that some ASIC technologies may be able to be used which would substantially reduce the cable-plant and result e.g. in much lower noise performance enabling better science.

### 4.2.2 Belle II

Except for the electromagnetic calorimeter, all major subdetector systems are being upgraded in going from Belle to Belle II. To realize optimal subdetector performance in the face of significantly increased event rates and backgrounds, each of these subdetector systems (pixel, silicon tracker, small-cell drift chamber, barrel and forward endcap particle identification, and muon system) upgrades involves at least one ASIC. Compared with their LHC counterparts, the radiation tolerance requirements are significantly relaxed.

### 4.2.3 LHCb

The key concept underlying the LHCb upgrade is the combination of new front end electronics to push out the data in "real time" without a hardware trigger, and an off-detector software trigger allowing event filtering with a software algorithm. Thus fast analog shaping, fast digitization and zero suppression and high data transmission bandwidth are key design goals in all the front end devices.

The LHCb tracker upgrade faces many of the same challenges faced by ATLAS and CMS. Radiation tolerance is one of them, for example the pixel detector front-end electronics is required to withstand about 400 Mrad in a 10 year timeframe. The LHCb collaboration has not yet decided on whether to upgrade the VELO using hybrid silicon pixel detectors or silicon strip detectors. New ASICs are being designed for both options. They all involve fast analog front end, digitization implemented either with the time-over-threshold method (VELOPIX) or analog-to-digital conversion method (SALT), zero suppression, buffer, and serializer to transmit the data from the detector.

In addition, a large effort is put in photon detector readout from a variety of devices. ASIC are being developed to solve problems of reduced impedance and faster shaping time, fast TDC or ADC digitization, single photon counting with Ma-PMTs (and MCPs and SiPMs), lower noise and fast return to the baseline for lower gain photomultiplier tube operation, and of course fast readout (see ASIC table in Appendix).

All these ASIC are being developed and are at different levels of maturity. In addition to the front end ASIC, an area of common interest is the transition to optical, implemented either in the front end or in a stage just outside the detector (PIXEL/Strips). Currently the CERN GBT ASIC is used. A lower power solution would be of great interest. Finally radiation tolerant DC-DC converters or linear regulators are items that will be part of the power distribution system.

### 4.2.4 Mu2e

The design of the Mu2e experiment allows the exclusive use of commercial electronics. However, the mechanical constraint posed by limited space between the outer edge of the active tracking volume and the solenoid magnet motivates the use of an ASIC. The experimenters currently plan to read out the straw tracker using commercial preamplifiers mounted on each end of the straws together with an ASIC containing a post-amplifier/pulse shaper, ADC, and TDC. A full ASIC solution is also being investigated. The ASICs will be mounted at the outer edge of the tracker midway between the two ends of the straws.

#### 4.2.5 Project X

Project X is a proposal to increase the proton beam power available at FNAL in a number of stages. Each stage will enable new experiments requiring high intensity. These experiments will require radiation tolerant, high rate precision calorimetry and low mass high rate charged particle tracking. Radiation tolerant ASIC front end electronics will be critical, and ASICs will also likely be required to achieve the necessary timing resolution.

### 4.3 Cosmic frontier

At least three cosmic frontier areas could require ASIC development in the future: waveform digitizers for large area Cherenkov UHE gamma and neutrino detectors; clock, bias and signal processing for astrophysics imaging and spectroscopy; and integrated RF frequency or time de-multiplexing components for large pixel count CMB focal planes.

For waveform digitizers, required characteristics are analog sampling memories with GHz input bandwidth, GSamples/sec sampling rates, and millisecond storage depth. Low cost per channel, a few $10s, and low power, 10–50 mW/channel, are required for high pixilation cameras. The TARGET and DRS4 ASICs (see ASIC table in the appendix) are today's state of the art.

ASIC development will be needed for CCD readout to support ultra-large pixel count focal planes or to reduce instrument weight and heat dissipation. For the associated large telescopes with their short exposure times, there will be continued pressure to reduce CCD readout time while not compromising read noise performance. One approach is CCDs with a large number of output ports each operating at modest pixel rates. ASICs will be required to minimize the power for the large number of analog processing chains and to provide the various bias voltages. See the ASIC table in the appendix for chip sets developed for LSST and JDEM that provide complete CCD control.

CMB focal planes with a million pixels are being discussed. Today's superconducting focal planes using TES or MKIDS have hand crafted electronics. The future will require integrated solutions of RF frequency de-multiplexing or time division multiplexing. Early phases of this today are integrated inductor arrays for frequency de-multiplexing. It will need to be explored how ASICs could be utilized especially for space applications.

ASICs have also been an enabling technology in sub-orbital and terrestrial searches for cosmogenic neutrinos, the existence of which has been predicted since the 1960s, though none have yet been measured. As it has now been demonstrated that Teraton scale detectors are required for such an observation, future, high-performance ASICs will be essential for improvements to the discovery sensitivity of these experiments.

Another area is cold cosmic frontier electronics (e.g. DarkSide) which is covered in the Intensity frontier LBNE section.

## 5 Future IC's: R&D needs

R&D is required in several areas in order to advance the functionality and performance of ASICs so as to improve or even enable future experiments. Technical challenges arise when the performance needs to be significantly improved or when operating conditions lie well outside of industrial applications. The latter presents a problem for modern ASIC design, which relies heavily on accurate simulation models. In the following sections the most relevant areas of R&D are summarized with more detailed explanations listed in the appendix for some of the topics. Additionally, it is expected that new areas of R&D will be added in the future which depend on instrumentation needs for experiments not yet proposed.

## 5.1 High-bandwidth transmission

Next generation detector systems require transmission of large data volumes from the detector. In-detector data transmission ASICs are needed either for high radiation, low temperature, or low background requirements, or where space or interconnectivity require the integration of high speed transmitter blocks with other functions. In some applications the availability of higher speed transmitters can reduce the need of lossy less flexible data reduction inside the detector volume which can degrade the physics performance. For receiving optical control and timing signals inside the detector volume, photo-diodes, trans-impedance amplifiers and deserializers, together with a transmission protocol that allows for bit error detection and correction are required. (See more in the appendix)

Especially for high-rate collider detectors, not all the data can be sent out of the detector, thus in-detector data processing is needed as described in the next section.

## 5.2 In-detector digitization, data reduction, processing

For experiments where in-detector data reduction is a viable or required solution, the front-end power and the number of interconnections to the outside of the detector can be reduced. Integrating the analog circuits with trigger logic and event data pipelines, and/or ADCs (Analog-to-Digital-Converters), or DSPs (Digital Signal Processors) or other processing blocks allows improved performance, power and cost reduction, and system optimization. (See more in the appendix.)

## 5.3 Radiation tolerance

Radiation tolerance R&D is driven by the needs of the LHC experiments, especially for the inner tracker layers. The current plan for the future is to use a commercial 65-nm CMOS process where CERN is leading the effort to validate the process and coordinate the generation of a design library.

A CERN RD collaboration is being formed to address the 1 Grad tolerance needed for the future innermost layers of ATLAS and CMS (RD53 Collaboration, "RD Collaboration Proposal: Development of pixel readout integrated circuits for extreme rate and radiation ," CERN-LHCC-2013-008, 2013). In the long term, even smaller feature sizes than 65nm will be desirable, but R&D is needed since the radiation tolerance of those processes is not known. In addition to CMOS, other technologies, e.g. germanium doped silicon (SiGe) bipolar technology, are being explored. (See more in the appendix,)

## 5.4 Low-temperature

The design of front-end CMOS ASICs operating in cryogenic (mainly liquid Argon ~ 70K and liquid Xenon ~ 170K) environment poses several challenges. Transistor models provided by vendors are not valid at those temperatures, so accurate models (static, dynamic, noise, and lifetime in strong, moderate and weak inversion) need to be obtained from measurements and extraction. (See more in the appendix.)

## 5.5 Low radioactivity

For some experiments low radioactivity is a requirement. Low levels (< $10^{-6}$ ppm) of elements like U, Th, K40 are especially important for underground experiments (e.g. nEXO, Darkside). Integrated circuit processes proposed for such experiments need to be investigated for suitability, e.g. via Inductively Coupled Plasma Mass Spectroscopy (ICPMS) or Neutron Activation (NAA).

## 5.6 Non-standard processing

Standard integrated circuits include many metal interconnect layers, but only one layer of transistors. 3D technologies allow the formation of ICs with more than one transistor layer. This allows many more transistors to be physically close to one another (meaning lower capacitance interconnects) than in a 2D circuit. True 3D circuits are not yet generally available, and it is not clear which of the many competing enabling technologies will become commercially viable, but some of the key technologies are well established. These include the formation of through-holes using Deep Reactive Ion Etching (DRIE) and wafer thinning by a combination of grinding, etching, and Chemical Mechanical Polishing (CMP). HEP designers are already beginning to use these technologies in "2.5D" designs that do not use more than one transistor layer, but have other advantages. One example is circuits in which through-holes allow bonds to be made through the chip from the backside of the circuit. Another is the use of silicon interposers with bump bonds on both sides connected by through-hole vias.

Two other examples of non-standard processing may be used soon for Monolithic Active Pixel Sensors (MAPS). One is to develop a quadruple well 180nm bulk CMOS process with a thick epitaxial layer, and another is to develop a modified Silicon-On-Insulator (SOI) process including a nested well structure.

## 5.7 System-on-chip

Modern integrated circuit technologies allow us to aggregate on the same chip several functions traditionally relegated to separate components. This is what system on chip (SOC) means. Work is on-going to integrate analog and digital signal processing, power regulation, monitoring, and safety interlock functions. For applications where silicon is the sensing material, there have been decade-long efforts to try to integrate sensor, amplification, and digital processing all in the same chip, as in the case of MAPS. More recently there has also been interest in integrating photon detectors such as Silicon Photomultipliers (SiPMs) in the same substrate.

3-D integrated processing offers the widest range of system integration options, but by the same token the type and number of layers to be vertically integrated must be chosen to suit the system needs. (See also the previous section "Non-standard processing"). In some cases, it may only be possible to solve a problem using a SOC approach. One example is an associative memory for fast reconstruction of tracking detector events. (See more in the appendix.)

## 5.8 High dynamic range

A key figure of merit in front-end electronics is the dynamic range, defined as the ratio between the maximum and the minimum measurable charge. In most practical cases the dynamic range is limited by the processing circuits that follow the analog front-end, such as discriminators and peak detectors. A major challenge with deep submicron technologies comes from the decreased supply voltage, now approaching ≈ 1V. In order to achieve a dynamic range in the ballpark of a few thousand without substantial increase in area and/or power, rail-to-rail and low-noise filter design techniques must be adopted. Due to the unique features of filters for radiation detection, it is expected that such new design techniques will require R&D effort. (See more in the appendix.)

## 5.9 Fast timing

ASIC technologies offer the ability to provide both time and charge information for use both "off detector" and "in-situ".

Time as a measured quantity (Off Detector use): The measurement of the time of arrival of a sensor signal relative to a reference clock requires a good match between two basic domains: the analog signal processing and the time measurement domain. For signals with fixed shapes, time invariant techniques such as constant fraction or zero crossing have already been implemented in ASICs. More sophisticated techniques that aggregate information from multiple channels are possible as well.

Fast timing for sensor coincidence tagging (In situ use): Complex, high density detector systems can benefit immensely from low latency event data filtering based on intelligent information constructed locally in sub-detectors. For the LHC upgrade several ideas have been proposed to create vector quantities within local areas of tracking sub systems describing track segments based on multiple measurements of tracks in R, phi, eta and Z. New sensors under development place much tighter timing resolution constraint on the electronics, the 4D Ultra-Fast Silicon Detector (4D-UFSD) will require timing resolution of ~10 ps to accomplish 10 μm spatial resolution.

In analog waveform sampling ASICs input waveforms are sampled at multi Gbit rates typically via delay-lock loop timing circuits. Trade-off is generally the number of storage cells for each channel and the maximum analog input bandwidth achievable. Dynamic range is limited by the maximum supply voltage and the size of the sampling capacitor (kT /C noise), which in turn limits the maximum input bandwidth. Sampling rates increase with smaller feature sizes or faster processes. Waveform sampling ASICs are used e.g. for cosmic frontier experiments.

It is clear that there are more advances possible in particle physics experimentation as well as other areas (e.g. PET and pCT - proton computed tomography - medical applications) with the improvements in timing precision possible with newer IC technologies provided necessary development is funded.

### *5.10 Reliability*

Solid state or semiconductor electronics is known for its high reliability. This expected longevity of semiconductor devices is especially important for most particle physics experiments targeted for 10 year+ run time. Furthermore, access to the electronics in these modern detectors is quite limited or impossible. There is evidence that the some devices in the newest IC technologies may not follow the traditional bathtub curve of failure rate[1]. Rather than the failure rate remaining flat for thousands of hours before rising abruptly at wear out, some wear out mechanisms engage much earlier resulting in a slow rise in failure rates over the entire expected lifetime of the components. The causes of this wear out are not new but may need accurate modeling for use in HEP. They include electro-migration, hot carrier injection, time dependent dielectric breakdown and negative bias temperature instability.

Another area of concern is operation at cryogenic temperatures (~ 70K), well below the minimum temperature evaluated and guaranteed by CMOS foundries (233K). Most of the major failure mechanisms, such as electro-migration, stress migration, time-dependent dielectric breakdown, and thermal cycling, are strongly temperature dependent and become negligible at cryogenic temperature. The remaining mechanism that can substantially affect the lifetime of CMOS devices due to aging is the degradation due to impact ionization, which causes interface state generation and oxide trapped charge.

R&D will be needed to establish adequate design criteria to achieve long term high reliability in these technologies.

[1]Mark White & Joseph B. Bernstein, "Microelectronics Reliability: Physics-of-Failure Based Modeling and Lifetime Evaluation", JPL Publication 08-5 2/08, http://trs-new.jpl.nasa.gov/dspace/bitstream/2014/40791/1/08-05.pdf

## 6 Findings and Recommendations

### *6.1 Findings*

Below findings from the workshop are listed:

1) Use of ASICs is often critical to enable an experiment, but even for experiments that could be done without ASICs, use of ASICs generally leads to improved performance and reliability. ASICs will be necessary for essentially all detector subsystems at the HL-LHC. Most or all intensity frontier experiments will need ASICs, even if the needs of some experiments are not yet well developed. Even ground based cosmic frontier experiments will need ASICs to manage ever larger channel counts and meet several other requirements including performance.

2) It was recognized that the science enabled by IC developments has been impressive. Yet most of these developments have been incremental (not surprisingly as in the microelectronics industry). The analogy was made that most baseball games are won by lots of singles, few by home runs. Too much funding emphasis on "home runs" at the expense of "singles" is detrimental.
3) Close communication between physicists and IC designers is essential for successful development of new IC's. Developing IC's from specifications, without interaction leading to optimization, does not work.
4) HEP has spearheaded the use of ASICs, but there is a growing need and adoption by other disciplines- not only Nuclear Physics which has a close connection to HEP. The main experience base is currently in HEP and increased application of this experience to other disciplines is of mutual benefit, as it allows IC groups to function efficiently.
5) ASIC design capability in the US HEP community is not concentrated in one location, but rather spread over several laboratories and universities. It is essential because it facilitates the necessary intimate connection between detector designers and ASIC designers. This is especially important for front-end electronics, where the creative tension between the desirable, the possible, and the affordable is a key element of the design process.
6) Training of engineers and physicists at universities for HEP should include an understanding of the capabilities and limitations of ASICs given their importance in modern experiments.
7) R&D is needed to evaluate new technologies for their suitability for HEP, to develop new device structures, and to improve the performance for future experiments.
8) It is important to keep pace with industry technology development. However, more modern processes are increasingly more complex. The design manual for 65nm CMOS is more than 3 times the size of the manual for 130nm. Mastering new processes demands increasingly more effort from IC design groups.
9) Most ASICs in HEP are currently designed in 250nm and 130nm technology. It seems that 250nm technology will be offered for at least 10 more years, and 130nm technology even longer. Some applications benefit from smaller feature sizes.
10) Multi-project fabrication services as provided by MOSIS (US) or Europractice (Europe) are essential to substantially reduce the cost of prototype circuits.
11) CERN provides prototyping access to the world-wide HEP community (including the US) for specific processes suitable for the radiation environment at LHC. CERN plans for future support continue to be generously inclusive.
12) Finding ways to balance designer work load as projects end and new ones begin was a serious concern of all groups, both labs and universities. All groups develop ASICs for non-HEP research, such as BES, and this provides some level of load balancing, but may not be enough especially for smaller groups.
13) The main barrier against a designer at institute A working a small fraction of time on a project from institute B seems to be bureaucracy preventing institute B from paying for design time at institute A in a simple way. For large projects, funding agreements or work for others contracts are set up, but the paperwork involved is too large and too slow for small jobs and the overhead costs are prohibitive for smaller groups including universities.
14) ASIC design in a European country (France) was discussed and the recommendations include items which might be positive to pursue in the US.

## *6.2 Recommendations*

1. Continue to encourage the strong physicist-IC designer links in the US. This is a vital part of innovation and also important to the educational/training mission.
2. Seek to increase generic ASIC R&D to keep up with technology.
3. Basic literacy on IC technology should be included in the education of physics students to facilitate the communication between physicists and engineers, which is especially true for analog circuits for detectors.
4. To facilitate communication among designers, hold a yearly workshop of US IC designers. Include technical training to keep up with industry developments.
5. A point-of-contact for each institute should be identified to facilitate communication between groups and to follow up on recommendations in this report.
6. Investigate practical options for a designer at institute A to work a small fraction of time on a project at institute B on which institute A is not involved. This would be very helpful for load balancing in small groups- particularly universities.
7. Complete and maintain an up-to-date catalog of existing ASICs as shown in the appendix 6.4.
8. Consider a scientific ASIC design stewardship role for HEP, analogous to the particle accelerator stewardship role.

# 7 Appendix

## 7.1 Technology overview

### 7.1.1 Motivation for types of IC fabrication processes and pros/cons

Most IC's for HEP applications currently under development are using one of the CMOS technologies with feature sizes of 65nm, 130nm, 180nm, or 250nm. No 65nm or 130nm chips can be found in currently operating detectors. The first 130nm chips will start taking data within 2 years.

Each of these 4 nodes has specific features, along with development and production costs that increase with decreasing feature size. The most expensive and newest node used for HEP design projects, 65nm, provides the highest density and speed, but no better analog performance and worse dynamic range than the larger feature size processes. It is therefore expected that the 130nm to 250nm nodes will continue to be used for as long as they are available, for applications where analog performance is key, where density is not a driving factor, and where cost is important (e.g. currently a production run in 250nm is ~$100k whereas for 65nm it is ~$900k). There may be other unique features that make one node best for a specific purpose. (See radiation tolerance and cryogenic sections.)

It is not simply a feature size that is commonly used, but a specific process from a specific vendor. There are three reasons for this. First, it is often necessary to qualify a fabrication process for the intended use (for example radiation tolerance), and it is best to use a process that has already been qualified. The second is that one can save design overhead by re-using or re-adapting an IP (Intellectual Property) block from a previous project. In addition it makes good management sense to limit the number of technologies our design teams need to know from an economical, schedule and risk vantage point. Therefore, once a process is used, it quickly becomes the default one to consider first for new designs. The third consideration is the frequency of the multi-project fabrication runs. For the case of LHC experiments standardization comes about through long-term frame contracts negotiated by the CERN microelectronics group with specific foundries for specific processes.

### 7.1.2 Present and future IC processes

Advances in particle physics research have been greatly enhanced by the availability of highly advanced integrated circuit technologies. Foundry operations provided by many IC fabrication facilities enable the physics community to design ICs specific to the needs of a particular experiment thus optimizing the performance for the experiment. While some of these design requirements have a commonality with other applications, very few match those of a commercial market. Modern day particle physics experiments would not be possible without the custom design ICs they employ.

Most of the foundry operations are accessible through the MOSIS organization in the U.S and the Europractice organization in Europe. These two organizations are particularly helpful because they combine IC designs from many institutes into one foundry fabrication run such that the fabrication costs are shared by several projects. These multi-project fabrications are extremely important for prototyping designs. Most foundries will also deal directly with customers but usually without this multi-project option although some do have their own multi-project offerings. In the past few years, during the development of the LHC and its experiments, CERN related IC designs have been numerous enough that the CERN micro-electronics group has provided a service to merge many designs into multi-project fabrication submissions and provide technical support for a limited number of foundry processes. There is a small cost advantage versus the MOSIS service with the main advantage being that there is no minimum size (i.e. cost) for a design project, so small test circuits can be fabricated for less money. It is not guaranteed but likely that this service will continue in the future.

The IC processes available cover a wide range of technologies and generations of development. These include CMOS, BiCMOS, bulk silicon, silicon-on-insulator, and silicon-germanium (SiGe). Some older generations, e.g. 350 nm and 250 nm feature sizes, are still available while newer, higher performance generations, e.g. 130 nm, 65 nm and 45 nm, are not only available but becoming the standard.

The more advanced technologies offer several advantages. They allow higher speeds and usually lower power, both of which can be important for large experiments requiring millions of channels or special purpose applications requiring very precise timing. The smaller feature sizes also make possible smaller segmentation of detectors, e.g. silicon pixel detectors, affording better position resolution or lower occupancies. There are, however, downsides to these cutting edge technologies. They are always more expensive and require significantly more resources for training and sophistication in the checking required before submission. While the increase in cost/mm is somewhat mitigated by the reduction in area required for specific circuits, the entry costs (e.g. mask costs) for prototyping can become quite expensive. Also, the reduction in rail voltages becomes an issue for both I/O and analog circuits. The latter are essential components of a detector readout system and are becoming more challenging as the supply (rail) voltage approach 1 V.

For many experiments, older technologies may be quite adequate but they have a finite lifetime as foundries find few customers for the older, slower processes. 800 nm technologies are no longer available. It is not clear how long the 350nm will remain available. 250 nm and 130nm should be available for at least 10+ years. This trend of the industry as well as the advantages of new technologies for some experiments, forces the community to always be looking to consider new ones.

When a new technology is considered for a project, many of its characteristics must be evaluated. Radiation hardness is the most common as it is a requirement for several physics experiments but is not evaluated by the foundry developers. Deep sub micron processes down to 65nm have transistors that are well characterized for their analog performance, however, in some cases noise and dynamic range have been harder to predict in simulations. Shrinking supply voltage and increasing leakage currents pose important new problems for analog designs in smaller geometry technologies. The importance of the small feature size technologies is in their capacity to provide complex digital functionality along with the analog processing required. Along with the benefit comes a much higher level of complexity. Increasing constraints posed by mixed mode design flows and increasingly complex design rules that require the use of more than one checking program in these technologies requires a high degree of specialization for each new technology.

While the benefit for particle physics is huge, it is important to keep in mind that accepting a new technology into the HEP repertory requires considerable amount of work including designing, fabricating and testing parts. This should be understood to be part of the R&D overhead of advancing detector electronics.

## 7.2 Design: Tools and methodology

### 7.2.1 Tools & design support

Since about 1990, integrated circuit technology has been increasingly used for analog signal processing, data acquisition and more recently for triggering. During that time both the technology and CAD tools required for design and verification have becoming increasingly sophisticated. Today two design platforms, Cadence and Mentor Graphics, have emerged with widespread foundry support for analog, digital or mixed signal designs. A lower cost tool, Tanner EDA, is also used for design and verification steps for several HEP and BES ASICs. For primarily digital designs, defined using a high level description language, Synopsys offers a complete design suite. These design platforms integrate countless specialized tools and hold database information that defines the ASIC designs from first simulations through to the mask description files sent to the foundry.

The use of commercial Intellectual Property circuit blocks within HEP ASICs should be explored more but the legal aspect needs to be investigated.

**Project level requirements:** As explained elsewhere, fine lithography processes enhance our ability to aggregate functions and increase the number of elements serviced by a single ASIC, which is great for physics reach, but they also introduce new layers of complication with which US researchers must be familiar in order to be competitive on an international scale. An extreme example, perhaps the most complex for HEP today, the ATLAS FEI4 PIXEL chip that uses 90M transistors to service ~27k pixels was designed by a team of specialists from five international institutes (LBNL from the US). It required interoperability of the CAD tools among participating groups and an additional layer of database management, SOS by Cliosoft, unfamiliar to most HEP groups, to synchronize the design among all institutes (this is the analog of an SVN repository for software development). However, for most ASICs on the horizon for non-LHC, non-inner-layer projects the overhead of multi-institution collaboration may not be justified and it may remain more efficient to have it designed by a single institute.

**Collaboration thoughts**: The level of collaboration in ASIC design is increasing in some areas, as mentioned in several sections of this report. We expect to see more and more collaborative designs as it becomes necessary to pool expertise to deal with more advanced technologies. At the same time, the need for ASICs in future experiments is growing, while resources are limited. Additional ways for institutes to collaborate may offer ways to maximize productivity with the limited resources available.

Possible concepts for new forms of collaboration follow. Some of these had broad agreement and have already been given as recommendations. The list below also includes ideas that may have had limited support and should not be taken as recommendations.

- US IC for HEP designer workshops - to promote information, exchange and collaborations among US IC for HEP groups. E.g. Intellectual Property (IP) circuit blocks could be used and shared between institutes.
- US HEP IC database - to provide up-to-date list of designs, with basic and contact information for each (an example is the table in this paper with additional information); may include non-US designs, technologies used.
- Identification of a preferred list of processes for projects requiring special qualifications (radiation, cryogenics).
- Sharing of qualification tasks. This can be in coordination with non-US institutes (e.g. CERN) or where design and verification technology knowledge facilitates adoption of new difficult technologies.

## 7.3 Future IC's: Challenges (in more detail)

In the following sections some of the areas discussed in the main body of this report are repeated with additional information included.

### 7.3.1 High-bandwidth transmission (radiation tolerant)

Next generation detector systems, especially for LHC or ILC, require transmission of large data volumes from the detector. Constraints are power dissipation, space, reliability and for some sub-systems also radiation tolerance or operation at non-standard temperatures. High speed links are also used for applications where streaming data lossless out of the detector with off-detector triggering and filtering is preferred over lossy in-detector data reduction. Advantages can be more flexible processing and the availability of data from more channels or sub-systems over longer time spans with increased processing power and memory banks. As an example more complex pick-up/common-mode noise corrections can be made off-detector before threshold cuts are performed. Disadvantages are higher number of data links and feed-throughs. Mitigation is the development of higher density feed-throughs.

Optical transmission is preferred because of its high bandwidth and long distance transmission features and the elimination of ground-loops in the communication path. ASICs are needed to serialize the in-detector data and drive laser diodes where commercial devices cannot be used because of e.g. radiation requirements. Currently research is focused on links between 5 and 10 Gb/sec. Another area of research is optical modulators to achieve higher bandwidth above 10 Gb/sec. For receiving optical control and timing signals inside the detector volume, photo-diodes, trans-impedance amplifiers and de-serializers, together with a transmission protocol that allows for bit error detection and correction are required.

Especially for high-rate collider detectors, not all the data can be sent out of the detector, thus in-detector data processing is needed as described in the next section.

### 7.3.2 In-detector digitization, data reduction, processing

High rate collider detector sub-systems require significant resources: power, material, bandwidth and infrastructure to transmit data off the detector at beam crossing rates. This data may be the number of hits in tracking detectors or ADC (12-17bit) values from each segment in a fine granularity detector. Once off the detector, the data related to a beam crossing ends up being used in first level physics analysis .1 to 1% percent of the time. Since it takes significant energy and material to transmit data from the front end, where it is already stored in digital format, to a remote location, the most efficient technique is to store as much data on detector as possible until the decision what data to transmit to the data acquisition system is made. Furthermore, inactive material in the tracking volume from cabling, cooling and mechanical support structures interferes in a non correctable way with the quality of the data through multiple scattering and interactions. Assuming that the off-detector throughput rate is constant, it is clear that reducing the data flow from the detector by a factor of hundreds will have significant impact on the amount of power and material devoted to readout.

Complex, high density detector systems can benefit immensely from low latency event data filtering based on intelligent information constructed locally in sub-detectors. For the LHC upgrade several ideas have been proposed to create vector quantities within local areas of tracking sub systems describing track segments based on multiple measurements of tracks in R, phi, eta and Z. One case, Ingrid, proposes to utilize multiple measurements of charged particle tracks in an inert gas. Pixelized ionization sensors along with the time of arrival of the signal are used to establish track projections within a single ASIC. In this case nanosecond timing resolution will yield drift coordinate position resolution of ~100μm consistent with the dimensions of the ionization clusters. In another, more conventional, case a coincidence between axially aligned inner and outer layers of silicon strip detectors will be detected using a correlator that compares beam crossing coincident hits with PT acceptance data downloaded from the central track processor. This system must report all interesting tracklet information once every beam crossing with minimal latency. Depending on rate, it appears possible to create dead timeless tracking layers where beam synchronous tracklets from up to 30k silicon strips can be reported on a single fiber.

Analog-to-Digital Converters (ADCs) provide digitization of information (e.g. amplitude, timing) generated by analog front-end ASICs. As previously discussed, for most of the HEP experiments for LHC, the analog information can be stored in analog memories and multiplexed to ADCs, which can be conveniently located far from critical areas (i.e. areas with power, space, and radiation constraints). Commercial ADCs can be used in many cases and are already available in a broad selection to cover a wide range of resolutions, speed, and power.

However, the integration of moderate-speed (few MS/s) ADCs in front-end ASICs enable on-chip Digital Signal Processing (DSP), which means for example self-calibration, smart digital triggering, zero suppression, data compression, deep digital memories, fully digital communication (including chip-to-chip for complex triggering schemes). On-chip DSPs could result in a dramatic reduction in the complexity and bandwidth of the data acquisition system and could potentially enable new science. The integration of high-speed (tens to hundreds MS/s) ADCs enables the use of optimized on-chip digital filtering, while analog circuits can be limited to the (always essential) low-noise charge amplification and anti-aliasing filtering stages.

ADCs are already part of several front-end ASICs for future experiments (e.g. SiD KPIX) but further R&D is needed for various ADC architectures - either in voltage mode (mainly charge redistribution) or in current mode - ranging from successive approximation (SAR) to pipeline, flash, and clock-less. The main challenges come from the severe power constraints and from the coexistence with the high-precision low-noise analog circuitry.

### 7.3.3 Radiation tolerance

Several IC design challenges for HEP are in the category of operating conditions not covered by the device models of standard IC manufacturers. Even if the transistors of a process remain operable under these conditions, their characteristics may change, and if there is no model for the changed characteristics then modern IC design is not possible. This problem can be addressed in three ways: (1) using a special manufacturer that supports the desired conditions, (2) qualifying a standard process to certify that the models provided are still valid under the desired conditions, (3) developing custom devices and models.

Radiation tolerance is divided into two areas, total ionizing dose (TID) along with total non-ionizing fluence (e.g. from neutrons or other hadrons) and single event upset (SEU) tolerance. The highest dose and SEU tolerance near-term requirements are for the inner layers of LHC detectors, followed by different levels and different balance between total dose and SEU depending on experiment.

For TID and total fluence tolerance three methods have been used. In the 1990's IC's for collider vertex detectors were made with military foundries (method 1) that offered proprietary radiation hard CMOS processes with good results up to 10 Mrad (CMOS technologies are essentially immune to non-ionizing damage). This transitioned to method 2 in the last decade thanks largely to work by CERN to develop custom design rules for a commercial 250nm process. IC's designed with this method were hard up to 50 Mrad. Currently, a commercial 130nm process is being widely used after having been qualified (again largely by CERN) to 200 Mrad (method 3). However, experience with IC's made with this process indicated it is radiation hard well beyond this level, possibly up to 1Grad. The key to such radiation tolerance is the use of very thin silicon oxide layers necessary to achieve the 130nm feature size. At a thickness of a few nm, $SiO_2$ is no longer a good insulator due to quantum mechanical tunneling by free electrons. This prevents the buildup of trapped charge from exposure to radiation, which is the main way CMOS transistor properties are altered by radiation in larger feature size processes.

Being able to use method 3 for total dose tolerance is an ideal situation. This enables IC design using sophisticated commercial tools that rely on high precision device models. At the same time the level of validation needed increases. Small effects due to radiation gain in importance and thus a higher precision of the device models is required. The main task for future radiation hard IC design is to validate the new processes, hoping that method 3 can continue to be used. This includes validation of single transistors as well as digital cell libraries. For the near future, a 65nm feature size process has gained consensus for LHC applications. CERN is negotiating a frame contract with a foundry for this process. Once this happens, additional validation work will need to be carried out, and as noted this will be more demanding than in the past. An R&D collaboration is being formed at CERN that will largely conduct the tests needed, rather than CERN alone doing the job. This R&D collaboration is for design of the next generation hybrid pixel readout chips for ATLAS and CMS, which is the application needing the highest radiation tolerance, specified as 1 Grad TID. It is important to note that it is a specific process from a specific vendor that is validated, and as the validation becomes more demanding, it is not practical to do this for multiple vendors. A CERN frame contract ensures long term access to a process, and so validation and frame contract go hand in hand. The US benefits greatly from this arrangement with little investment.

Continued application of method 3 for the longer term is not guaranteed. Good as they are for radiation tolerance, leaky oxide gates are unfortunately not ideal for transistor operation and standing power consumption.

Starting with the 45nm node, IC manufacturers began replacing the $SiO_2$ gate dielectric with much thicker high K insulators. The radiation tolerance of these processes has not yet been explored. First evaluations should be done within a few years in order to understand how radiation hard IC design might evolve. It could well be that the 65nm process mentioned above marks the end of a heyday in radiation tolerant design.

For some detector applications, mostly analog, technologies other than CMOS offer performance advantages. However bipolar technologies are not as immune to ionizing radiation as most advanced CMOS technologies and also suffer from non-ionizing damage. They can nevertheless be qualified for use in many applicable radiation environments. The present ATLAS detector makes use of a bipolar technology for two of its readout systems, which were made possible by method 1 above. It is unlikely that any future bipolar technology will be developed especially to be radiation hard. However, just as with CMOS, the smaller feature sizes of the newer commercial technologies are providing better levels of radiation immunity. A new germanium doped silicon (SiGe) bipolar technology looks promising for an upgrade of the readout of the ATLAS liquid argon calorimeter using method 3. Power regulators also commonly use bipolar devices, qualified for radiation environments by method 3. As with CMOS, these other technologies require considerable effort to evaluate their hardness against both TID and total fluence. SiGe technology is not only used to produce bipolar devices within a CMOS process, but also to produce strained lattice CMOS silicon transistors (in which case the Ge alloy serves purely a mechanical function), and such strained lattice silicon transistors are only affected by TID just as their plain CMOS counterparts.

SEU tolerance is an entirely separate consideration from total dose tolerance. The same small features sizes and thin oxide layers that result in high total dose tolerance translate into a low energy threshold for SEU. This is essentially the energy required to change the logic state of a gate. On the other hand the probability that an SEU occurs goes down since the area is smaller for a given cell. SEU tolerance is achieved mainly by circuit design and layout techniques. The transistors and logic library cells of a given technology must be characterized for SEU threshold and cross section, and the IC designers must then build enough redundancy and physical separation between redundant elements to meet SEU tolerance specifications. SEU tolerance is thus design-specific rather than technology-specific.

### 7.3.4 Low-temperature

The stringent requirements on low-noise, low-power, precise signal processing and, in most cases, long lifetime (in excess of 20 years with sufficient margin) is possible only if accurate CMOS cryogenic models are made available. Almost all of commercial CMOS vendors focus their models in temperature ranges from -40C to 125C and on device lifetimes of about 10 years. The design of cryogenic front-end ASICs for HEP requires models capable of accurately reproducing the static and dynamic response, the noise performance, and lifetime of CMOS devices and circuits operating down to the ~-200C/70k range. These models must extend down to the weak-to-moderate inversion region, considering the low-power requirements on analog circuits.

A controller ASIC for imaging photodiode arrays has been developed by a commercial company, Teledyne Imaging Systems, which operates down to -235C/35K and is commercially available.

A small number of HEP groups (BNL, FNAL in collaboration with SMU, LBNL) made an effort to develop models in support of ASIC designs for small and medium-size cryogenic detectors targeted for -200C/70k operation (MicroBooNE, LBNE, SNAP/JDEM). SLAC developed models for cryogenic liquid-xenon operation at -100C/170K (nEXO). But overall the results are partial, and only limited to a few CMOS technologies: 180nm, 130nm, 250nm, and 800nm (SOI).

There is a need for a more systematic characterization and modeling of CMOS technologies in view of their operation in cryogenic environments for HEP. The characterization should include device response (static, dynamic, noise, and lifetime in strong, moderate and weak inversion) and digital sub-circuit response.

### 7.3.5 System-on-chip

Modern integrated circuit technologies allow us to aggregate on the same chip several functions traditionally relegated to separate components. This is not an HEP concept-industry has already led the way in SOC, since every function that can be absorbed into one integrated circuit reduces the cost of a system. This is, for example, the reason digital imaging is now so ubiquitous.

In scientific applications reducing cost is not the only driver, the SOC approach can often increase performance. In applications where the readout electronics are detector-mounted, SOC offers significant advantages by minimizing mass, volume, power and development cost, while increasing the possible bandwidth between processing stages and the channel density. Already in present detectors the analog front end, digitization, and digital I/O functions have been combined in single ASICs. This highly efficient approach requires a combination of both analog and digital processing on the same substrate. In the future it may be possible to integrate more and more detector functions into one ASIC. Work is on-going to integrate power regulation, monitoring, and safety interlock functions. For applications where silicon is the sensing material, there have been decade-long efforts to try to integrate sensor, amplification, and digital processing all in the same chip, as in the case of Monolithic Active Pixels (MAPS). More recently there has also been interest in integrating photon detectors such as Silicon Photomultipliers (SiPMs) in the same substrate. With a narrow window between avalanche and breakdown, today's SiPM arrays must be tuned to accommodate the noisiest elements since the output is the logical OR of many individual cells. A SOC approach integrating readout and sensor voltage control would allow masking noisy cells and optimizing the operating potential of each cell. This would allow lower gain operation, reducing noise (dark counts) and recovery time. Over the next decade, these and other SOC activities will continue and expand.

The desire to integrate more functions onto a single chip often requires a process with higher complexity and options. Sensing functions, higher voltages (for example for power conversion), isolation features, etc., require special process features and/or non-standard substrate wafers. 3-D integrated processing offers the widest range of system integration options, but by the same token the type and number of layers to be vertically integrated must be chosen to suit the system needs. All these options and special processing can in the end result in both development and device cost increases over less integrated solutions based on standard processing only, but they are nevertheless pursued by HEP R&D because they can result in higher performance. In some cases, it may only be possible to solve a problem using a SOC approach. One example is an associative memory for fast reconstruction of tracking detector events. Just as commodity microprocessors had to transition from single-core to multi-core in order to continue increasing performance, similar considerations about speed and power of data transfers is true for associative memory devices. They must soon transition to multiple tiers in order to continue to increase pattern density. Envisioned 3-D associative memory chips are often referred to as "experiment on chip", because each would have a pattern recognition capacity that in the past decade required several racks of electronics containing thousands of earlier version associative memory chips.

Most SOC solutions in the next decade will likely be incremental rather than transformational. While incremental in nature, this approach nevertheless permits higher performance and therefore greater science reach than possible before, e.g. in terms of channel count or processing speed. On the other hand, there might be cases where a SOC will be transformational, i.e. it will expand the system scope of a detector to include functionality not previously possible. A good example of this is addition of local triggering capability to charged particle tracking detectors. This requires massively parallel data processing in local detector elements, only possible with a SOC solution.

### 7.3.6 High dynamic range

Dynamic range is usually not limited by the first charge amplification circuit. Continuous-feedback charge amplifiers can exceed a dynamic range of hundreds of thousands by using non-linear voltage response while maintaining a low size and linear charge amplification. The actual limit in DR comes from time-invariant linear filters (shapers) since a reduction in the charge amplification to accommodate more charge in the shaper results in an increased noise from the shaper itself. The theoretical maximum DR achievable by a front-end ASIC is roughly given by $Q_{max}/ENC \approx V_{DD}/sqrt(4kT/C)$, where $V_{max}$ is the maximum voltage swing and C is the amount of capacitance used in the shaper. For low-noise linear analog front-end this theoretical limit is in the ballpark of several thousands. In some cases a few times $10^4$ has been achieved.

In most practical cases the dynamic range is limited by the processing circuits that follow the analog front-end, such as discriminators and peak detectors. Voltage offsets, capacitive injection and comparator hysteresis associated with these circuits set the minimum detectable voltage to a few mV which, in turns, limits the DR to $V_{max}$/few-mV. A major challenge with deep submicron technologies comes from the decreased supply voltage $V_{DD}$, now approaching $\approx$ 1V. Time-invariant linear front-ends based on standard design techniques wouldn't be able to exceed DRs of a few hundred. In order to achieve DR in the ballpark of few thousand without a substantial increase in area and/or power, rail-to-rail voltage (where $V_{max}$ approaches $V_{DD}$) and low-noise (shaper) design techniques must be adopted. Due to the unique features of filters for radiation detection, it must be expected that such new design techniques will require some moderate R&D effort. Deep sub-micron technologies offer the option of using thick-oxide MOSFETs at any point in the design. Such devices are capable of operating at voltages about twice the nominal one and would allow designers to double the DR. The main drawbacks are the need of a second voltage supply, some decrease in performance (the minimum channel length is about twice the nominal one), and the reduced radiation tolerance due to the thicker oxide. For example, the 130nm node (2.5 nm oxide) offers also ~ 5 nm oxide thickness, which corresponds to the 250 nm node, with still a high radiation tolerance.

Some HEP experiments require a DR in excess of few thousand. These front-ends can only be realized by using either a continuous non-linear filter or a time-variant filter. In both cases the design challenges are considerable. A commonly adopted solution consists of splitting the analog chain into two or more parallel paths with different gains: as soon as one path approaches the saturation the next path with lower gain is engaged and so on. Design challenges come from the trade-off between the number of independent paths versus the complexity, real estate and power dissipation. In principle this technique would allow arbitrarily high DRs. For very high dynamic ranges the continuous-feedback in charge amplifiers may need to be replaced with a switched circuit where the feedback components – either increasingly large capacitors or active devices – are enabled in real time based on voltage levels. In some cases charge subtractions – either capacitive or with current sources – at the input node or a suitable internal node can be adopted. But these solutions pose severe challenges, especially in those cases where fast processing is needed or the time of arrival of the charge is not precisely known.

All in all the development of design techniques for both high and very-high dynamic ranges may require substantial R&D effort and must be carefully taken into account when estimating the development time of front-end ASICs for HEP.

### 7.3.7 Reliability

We have become accustomed to the famous "bathtub curve" plotting failure rates as a function of operating time, which typically shows a relatively high failure rate in the first few hours of operation, referred to as infant mortality, followed by a long period of thousands of hours with negligible failure rate until wear-out starts to occur with a rise in failure rate. This typical characteristic of semiconductor devices allows those with fabrication defects to be weeded out quickly by a relatively short "burn-in" test without significantly compromising the lifetime of good parts and affords the assurance of reliable operation for many years under normal operating conditions. Extensive reliability studies have also shown that expected device lifetimes can be determined by accelerated aging tests usually at elevated temperatures and possibly voltages.

This expected longevity of semiconductor devices is especially important for most particle physics experimentation. As the field investigates new phenomena, the experiments must search for more and more rare interactions. This commonly results in data collection periods extending over many years. The typical lifetime target for most detector systems now is 10 years of operation. Furthermore, access to the electronics in these modern detectors is quite limited. As an example, access to the inner pixel and silicon strip detectors of ATLAS and CMS requires on the order of a year's downtime of the LHC machine. Similarly, access to experimental equipment in satellites is normally impossible. For these reasons, electronics systems for particle physics experiments place high importance on reliability. There is evidence that some devices in the newest, most advanced IC technologies may not follow the traditional bathtub curve of failure rate. Rather than the failure rate remaining flat for thousands of hours before rising abruptly at wear out, some wear out mechanisms engage much earlier resulting in a slow rise in failure rates over the entire expected lifetime of the components.

The causes of this wear out are not new. They include electro-migration, hot carrier injection, time dependent dielectric breakdown and negative bias temperature instability. As the feature size of these technologies decreases and the performance (mostly speed) increases, these failure mechanisms become more prominent and their onset earlier in the life to the device.

For commercial applications, these shorter lifetimes may not be a problem. Manufacturers are continually building new features into their products providing strong incentives to replace older models. If particle physics wants to make use of the improved performance of the continually advancing IC technologies, their reliability is an important characteristic that must is evaluated. Some of these failure mechanisms can be mitigated by backing-off the design specifications posted by the foundries. As an example, electromigration has a strong dependence on current density and temperature. Many detectors operate at cold temperatures to reduce noise or leakage currents. Increasing the width of conductors can lower current density and lengthen the time for this type of wear out.

Another area of concern is operation at cryogenic temperatures. (~ -200C) well below the minimum evaluated and guaranteed by CMOS foundries (-40C). Most of the major failure mechanisms, such as electro-migration, stress migration, time-dependent dielectric breakdown, and thermal cycling, are strongly temperature dependent and become negligible at cryogenic temperature. The remaining mechanism that can substantially affect the lifetime of CMOS devices due to aging is the degradation due to impact ionization, which causes interface state generation and oxide trapped charge. When a CMOS device is operated at cryogenic temperatures, the amount of impact ionization at a given operating point increases, decreasing its lifetime. While some key properties of CMOS transistors (noise, gm/Id ratio, speed) improve at low temperature, the drain operating voltage has to be slightly reduced for equal lifetime, due to increased mobility and reduced carrier mean-free-path.

Models provided by foundries are limited and may not be sufficient for HEP applications. Rules of thumb may be adopted to increase the lifetime, for example by operating analog circuits at low current densities (i.e. low-power design) and digital circuits at reduced voltage and frequency. A systematic R&D program is needed in support of ASIC design for long lifetime (room temperature and cryogenic). The R&D program should include (a) device-physics-based design guidelines for reliability and (b) accelerated lifetime stress tests. CMOS and BiCMOS technology nodes (90nm and below) expected to be used in future HEP detectors should be investigated. Understanding these wear out mechanisms relative to the specifications of each technology and requirements of the experimental equipment has become very important and requires testing of actual parts as well as simulations. This evaluation work must be factored into the adoption of any new technology for particle physics research.

Due to the large cost and inaccessibility of ASICs for space missions, in addition to circuit techniques to increase reliability, the entire process of design, fabrication, and testing is required to be evaluated, monitored, and documented. That is not just to be able to detect potential issues which could cause failures during the lifetime of the mission, but also to be able to investigate and determine the cause of any failures occurring during testing or during the mission. The ASIC fabrication process as well as each lot has to be qualified for space use (e.g. accelerated lifetime, humidity, radiation, temperature cycling, temperature range, vibration, EMI testing). Some of the qualification steps are destructive (e.g. SEM). In addition each device to be used for the mission has to go thru several screening steps.

### *7.4 ASICs in HEP experiments*

The following table lists some ASICs in use or proposed. The list is not meant to be exhaustive but to provide an indication of the number of IC design for HEP. Where US institutions are involved, the name(s) of the institution(s) are called out in the table. The state column indicates whether the ASIC is in a running experiment (R), for an approved experiment (A), or a candidate for a proposed experiment (C).

| Experiment | Sub-system | Name | Description | * | Frontier | Institution | Type | Technology |
|---|---|---|---|---|---|---|---|---|
| ATLAS | pixel | FE-I3 | pixel front end chip | R | Energy | LBNL | mixed | 250nm CMOS |
| ATLAS | pixel | MCC | digital I/O | R | Energy | - | digital | 250nm CMOS |
| ATLAS | pixel | FE-I4 | pixel front end chip | A | | LBNL | mixed | 130nm CMOS |
| ABCD | strips | ABCD | strip front end chip | R | Energy | UCSC | mixed | 0.8μm DMILL |
| ATLAS | strips | ABCn | strip front end chip | C | Energy | UCSC,Penn | mixed | 250nm CMOS |
| ATLAS | strip+pixel | DORIC | laser diode receiver | R | Energy | OSU | mixed | 250nm CMOS |
| ATLAS | strip+pixel | VDC | VCSEL driver | R | Energy | OSU | mixed | 250nm CMOS |
| ATLAS | strip+pixel | BPM-12 | laser diode driver | R | Energy | - | mixed | 250nm CMOS |
| ATLAS | strip+pixel | DRX-12 | laser diode receiver | R | Energy | - | mixed | 250nm CMOS |
| ATLAS | upgrade | ABC-130 | strip front end chip | C | Energy | Penn, UCSC | mixed | 130nm CMOS |
| ATLAS | upgrade | HCC-130 | strip module control | C | Energy | Penn, UCSC | mixed | 130nm CMOS |
| ATLAS | upgrade | SPP | Serial Power & Protection | C | Energy | Penn | analog w/ basic digital | 130nm CMOS |
| ATLAS | FTK | AM | associative memory | A | Energy | - | custom digital | 65nm CMOS |
| ATLAS | TRT | ASDBLR | straw front end | R | Energy | Penn | analog | 0.8u DMILL |
| ATLAS | TRT | DTMROC | straw digitizer | R | Energy | Penn | digital | 250nm CMOS |
| ATLAS | LAr Calo | CLKFO | clock fanout | R | Energy | Nevis Labs, Columbia | digital | 250nm CMOS |
| ATLAS | LAr Calo | Gain Selector | analog Range Selection | R | Energy | Nevis Labs, Columbia | mixed | 250nm CMOS |
| ATLAS | LAr Calo | SCA Controller | analog mem Control | R | Energy | Nevis Labs, Columbia | digital | 250nm CMOS |
| ATLAS | LAr Calo | HAMAC-SCA | analog memory | R | Energy | Nevis Labs, Columbia | analog | 0.8u DMILL |
| ATLAS | LAr Calo | BiMUX | analog mux | R | Energy | ? | analog | 0.8u DMILL |
| ATLAS | LAr Calo | OpAmp | op amp | R | Energy | ? | analog | 0.8u DMILL |
| ATLAS | LAr Calo | DAC | 16-bit DAC | R | Energy | ? | mixed | 0.8u DMILL |
| ATLAS | LAr Calo | SPAC slave | control logic | R | Energy | - | digital | 0.8u DMILL |
| ATLAS | LAr Calo | Configuration | control logic | R | Energy | Nevis Labs, Columbia | digital | 0.8u DMILL |
| ATLAS | LAr Calo | SMUX | data mux | R | Energy | ? | digital | 0.8u DMILL |
| ATLAS | LAr Calo | Calibrator | digital logic | R | Energy | ? | digital | 0.8u DMILL |
| ATLAS | LAr Calo | LAPAS | analog | C | Energy | BNL,Penn | analog | 130nm SiGe BiCMOS |
| ATLAS | LAr Calo | Nevis-12 | 40MHz ADC | C | Energy | Columbia | mixed | 130nm CMOS |
| ATLAS | Muon Small Wheel | VMM | front-end | A | Energy | BNL | mixed | 130nm CMOS |
| ATLAS | Muon Small Wheel | TDS | digital logic/serialiser | A | Energy | U. Michigan | digital | 130nm CMOS |
| ATLAS | LAr Calo | LOCx2 | 5.12 Gbps serializer | C | Energy | SMU | mixed | 250nm SoS CMOS |
| ATLAS | LAr Calo | LOC-D | 5.12 Gbps VCSEL driver | C | Energy | SMU | mixed | 250nm SoS CMOS |
| ATLAS | Tile Calo | TileDMU | pipeline | R | Energy | | digital | 350nm CMOS |
| ATLAS | Muon MDT | ASD | amplifier/shaper | R | Energy | Harvard | analog | 0.5u CMOS |
| ATLAS | Muon MDT | AMT | TDC | R | Energy | | mixed | 350nm CMOS |
| ATLAS | Muon CSC | ASM1 | preamp | R | Energy | | analog | 0.5u CMOS |
| ATLAS | Muon CSC | ASM2 | MUX | R | Energy | | analog | 0.5u CMOS |
| ATLAS | Muon CSC | Clock driver | clock driver | R | Energy | | digital | 0.5u CMOS |
| ATLAS | Muon CSC | MAMAC-SCA | analog memory | R | Energy | | analog | 0.8u DMILL |
| ATLAS | Muon RPC | ASD | amplifier/shaper | R | Energy | | analog | GaAs |
| ATLAS | Muon RPC | CMA | coincidence matrix | R | Energy | | digital | 180mm CMOS |
| ATLAS | Muon TGC | ASD | amplifier/shaper | R | Energy | | analog | Bipo;ar |
| ATLAS | Muon TGC | HpT | trigger | R | Energy | | digital | 350nm CMOS |
| ATLAS | Muon TGC | PP | trigger | R | Energy | | digital | 350nm CMOS |
| ATLAS | Muon TGC | SLB | trigger | R | Energy | | digital | 350nm CMOS |
| ATLAS | Muon TGC | JRC | JTAG controller | R | Energy | | digital | 350nm CMOS |
| ATLAS | upgrade | BCC | communication | A | Energy | SLAC, LBNL | digital | 250nm CMOS |
| PHENIX (*) | strips | SVX4 | strip front end chip | R | Nuclear Physics | LBNL, FNAL | mixed | 250nm CMOS |
| PHENIX (*) | strips | FPHX | strip front end chip | R | Nuclear Physics | FNAL | mixed | 250nm CMOS |
| CLAS12 | strips | FSSR2 | strip front end chip | A | Nuclear Physics | FNAL | mixed | 250nm CMOS |
| CMS & Belle II | strips | APV25 | strip front end chip | R A | Energy/intensity | - | mixed | 250nm CMOS |
| CMS | pixel | PSI46 | pixel front end chip | R | Energy | - | mixed | 250nm CMOS |
| CMS | pixel | TBM05a | pixel readout control chip | R | Energy | Rutgers | mixed | 250nm CMOS |
| CMS | HCal HPDs | QIE8 | pseudo floating point ADC | R | Energy | FNAL | mixed | 800nm BiCMOS |
| CMS | Px, Tk, ECAL, RPC | CCU25 | FE control | R | Energy | - | digital | 250nm CMOS |
| CMS | Px, Tk | LVDSMUX3 | FE control | R | Energy | - | digital | 250nm CMOS |
| CMS | ECAL | LVDSMUX4P | FE control | R | Energy | - | digital | 250nm CMOS |
| CMS | Px, Tk, ECAL | PLL25 | FE control | R | Energy | - | digital | 250nm CMOS |
| CMS | All over the place | LVDSBUF | FE control | R | Energy | - | digital | 250nm CMOS |
| CMS | Px, Tk, ECAL | DCU25 | FE monitoring | R | Energy | - | digital | 250nm CMOS |
| CMS | Px, Tk, ECAL, RPC | Rx40 | CDR | R | Energy | - | mixed | 250nm CMOS |
| CMS | ECAL | MGPA | multi-gain preamp | R | Energy | - | analog | 250nm CMOS |
| CMS | ECAL | AD9042 | 4 channel ADC | R | Energy | - | mixed | 250nm CMOS |
| CMS | ECAL | CRT910T | LVDS to LVCMOS | R | Energy | - | analog | 250nm CMOS |
| CMS | ECAL | FENIX | front-end Trig & DAQ | R | Energy | - | analog | 250nm CMOS |
| CMS | All over the place | GOL | serializer & LED driver | R | Energy | - | analog | 250nm CMOS |
| CMS | All over the place | QPLL25 | clock cleaner | R | Energy | - | analog | 250nm CMOS |
| CMS | All over the place | GOL | serializer & LED driver | R | Energy | - | analog | 250nm CMOS |
| CMS, NOVA | ECAL | ADC41240 | ADC | R | Energy | FNAL | analog | 250nm CMOS |
| CMS | Tk | LLD | laser driver | R | Energy | - | mixed | 250nm CMOS |
| CMS | Pre-shower | K-Chip | | R | Energy | - | mixed | 250nm CMOS |
| CMS | Pre-shower | Pace | | R | Energy | - | mixed | 250nm CMOS |
| CMS | Tk alignment | | opamp | R | Energy | - | analog | 250nm CMOS |
| CMS | HCAL PMTs | QIE10 (PMT) | pseudo floating point ADC | C | Energy | FNAL | mixed | 250nm SiGe |
| CMS | HCAL SiPMs | QIE11 (SiPM) | pseudo floating point ADC | C | Energy | FNAL | mixed | 250nm SiGe |
| | | | | | | | | |
| SID | ECAL, TKR | KPIX | 1k channel amp/ADC/core | C | Energy | SLAC | mixed | 250nm CMOS |
| SID | FCAL | Bean | 32-ch amp/adc | C | Energy | SLAC | mixed | 180nm CMOS |
| FERMI | Calorimeter | GCFE | calorimeter front-end | R | Cosmic | SLAC | mixed | 500nm CMOS |
| FERMI | Calorimeter | GCRC | digital controller | R | Cosmic | SLAC, NRL | digital | 500nm CMOS |
| FERMI | Tracker | GTFE | strip front-end chip | R | Cosmic | UCSC, SLAC | mixed | 500nm CMOS |
| FERMI | Tracker | GTRC | strip digital controller | R | Cosmic | SLAC | digital | 500nm CMOS |
| FERMI | Anti-Coincidence | GAFE | PMT front-end | R | Cosmic | SLAC | analog | 500nm CMOS |
| FERMI | Anti-Coincidence | GARC | digital controller | R | Cosmic | SLAC, GSFC | digital | 500nm CMOS |
| FERMI | Tracker | GTCC | trigger-readout controller | R | Cosmic | SLAC | digital | 500nm CMOS |
| FERMI | Calorimeter | GCCC | trigger-readout controller | R | Cosmic | SLAC | digital | 500nm CMOS |
| FERMI | Event builder | GEEE | lvds-converter | R | Cosmic | SLAC | digital | 500nm CMOS |
| nEXO | TPC | nEXO-FE | front-end chip | C | Intensity | SLAC | analog | 180nm CMOS |
| nEXO | TPC | nEXO-ROC | digitizer/controller | C | Intensity | SLAC | mixed | 180nm CMOS |
| SeaQuest & g-2 | wire chambers | ASDQ | amplifier shaper discriminator with charge-dependent width | R A | Intensity | Penn | mixed | "SHPi" bipolar |
| MINOS | multianode phototubes | VA32 HDR11 | multianode PMT front end | R | Intensity | - | mixed | .8micron CMOS? |
| MINOS | multianode phototubes | QIE7b | pseudo floating point ADC | R | Intensity | FNAL | mixed | 3micron BiCMOS |
| MINERvA | APDs | TriP-t | VLPC front end/trigger pipeline | R | Intensity | FNAL | mixed | 250nm CMOS |
| NOvA | APDs | NOvAchip | APD front end chip | A | Intensity | FNAL | mixed | 250nm CMOS |

| Experiment | Subsystem | ASIC | Description | State | Frontier | Location | Type | Technology |
|---|---|---|---|---|---|---|---|---|
| Belle II | Muon System | TARGET6B | scin-strip/MPPC readout | A | Intensity | Hawaii | mixed | 250nm CMOS |
| Belle II | Particle ID (Time Of Propagation detector) | IRS3B | Cherenkov, ps-timing det. | A | Intensity | Hawaii | mixed | 250nm CMOS |
| ANITA3 | RF trigger | RITC2 | correlation 3-bit trigger | A | Cosmic | Hawaii | mixed | 130nm CMOS |
| ANITA3 | RF digitizer | LAB4B | impulsive radio recorder | A | Cosmic | Hawaii | mixed | 250nm CMOS |
| CTA | Camera trig/readout | TARGET5/7 | MPPC/MA-PMT readout | C | Cosmic | Hawaii | mixed | 250nm CMOS |
| ARA | Antenna digitizer | IRS2 | radio transient recorder | R | Cosmic | Hawaii | mixed | 250nm CMOS |
| ANITA/AURA | RF digitizer | LABRADOR3 | radio transient recorder | R | Cosmic | Hawaii | mixed | 250nm CMOS |
| Belle II | Pixel detector | SWITCHERB | HV row control for DEPFET | A | Intensity | - | mixed | 180nm HV CMOS |
| Belle II | Pixel detector | DCDB | multichannel ADC | A | Intensity | - | mixed | 180nm CMOS |
| Belle II | Pixel detector | DHP | data handling processor | A | Intensity | - | digital | 65nm CMOS |
| Belle II | Wire chamber | ASD-CDC | wire chamber amplifier shaper discriminator | A | Intensity | - | mixed | 800nm BiCMOS |
| Belle II | Aerogel RICH | SA03 | APD front end chip | A | Intensity | - | mixed | 350nm CMOS |
| Mu2e | Wire chamber | POM | postamp/shaper/ADC/TDC | A | Intensity | LBNL | mixed | 65nm CMOS |
| LHCb | Silicon strip detector | BEETLE | strip front end chip | R | Intensity | | mixed | 250nm CMOS |
| LHCb | Hybrid photo detector | LHCBPIX1 | pixel front end | R | Intensity | | mixed | 250nm CMOS |
| LHCb | Silicon strip detector | SALT | strip front end chip | A | Intensity | | mixed | 130nm CMOS |
| LHCb | RICH | CLARO or MAROC3 | multianode PMT front end | A | Intensity | - | mixed | 350mn BiCMOS or 350nm CMOS |
| LHCb | Scintillating Fiber tracker | Pacific | SiPMT readout | A | Intensity | | mixed | 130 nm CMOS |
| LHCb | ECal | ICECAL | phototube amp/track&hold | A | Intensity | - | analog | 350nm BiCMOS |
| All LHC upgrades | Data readout | GBT | Multi Gbps transciever | A | Energy & Intensity | | digital | 65nm CMOS? |
| LHCb | Pixel detector | Velopix | pixel front end | A | Intensity | | mixed | 130nm CMOS |
| LBNE | Anode Plane Array | LArFE | wires front-end | A | Intensity | BNL | analog | 180nm CMOS |
| LBNE | Anode Plane Array | LArADC | S&H, ADC, multiplexing | A | Intensity | BNL | mixed | 180nm CMOS |
| LBNE | Anode Plane Array | LArMUX | digital multiplexing | A | Intensity | FNAL or BNL | digital | 180nm CMOS? |
| LBNE | Anode Plane Array | LArREG | voltage regulator | A | Intensity | GeorgaTech | analog | 180nm BiCMOS? |
| IceCube | DOM | ATWD | PMT transient recorder | A | Cosmic | LBNL | analog | ? |
| KamLand | | ATWD | PMT transient recorder | A | Cosmic | LBNL | analog | |
| LSST | CCD Readout | CABAC | CCD Clocks And Biases | A | Cosmic | LPNHE | mixed | 350nm CMOS, HV |
| LSST | CCD Readout | ASPIC | CCD Analog Signal Processing | A | Cosmic | LAL, LPNHE | analog | 350nm CMOS |

Table 1: List of ASICs: * State column: In running experiment (R), for approved experiment (A), candidate for proposed experiment (C).